\documentclass[a4paper,natbib2]{jpconf}
\usepackage{graphicx}

\newcommand{\mnras}{MNRAS}
\newcommand{\aap}{A\&A}

\newcommand{\apjl}{ApJL}

\def\sgra{Sgr~A$^*$}

\def\msun{{\,{\rm M}_\odot}}

\def\ltsima{$\; \buildrel < \over \sim \;$}
\def\simlt{\lower.5ex\hbox{\ltsima}}
\def\gtsima{$\; \buildrel > \over \sim \;$}
\def\simgt{\lower.5ex\hbox{\gtsima}}

\begin{document}

\title{News from the year 2006 Galactic Centre workshop}

\author{Mark Morris$^1$ and Sergei Nayakshin$^2$}

\address{$^1$ Dept. of Physics \& Astronomy, Univ. of California, Los Angeles, USA \\
$^2$ Dept. of Physics \& Astronomy, University of Leicester, LE1 7RH,
  UK}

\begin{abstract}
We summarize some of the new results from contributions made to the
Galactic Centre workshop that took place in Bad Honnef, Germany, on
April 18-22 2006.
\end{abstract}

\section{Introduction}

One of the web definitions of ``a center'' reads: ``A place of concentrated
activity, influence, or importance''
(http://www.answers.com/topic/center).  Another adds, ``a point of origin from
which ideas or influences originate''.  We believe this conference on the
Galactic Center is a great illustration of these definitions for several
important fields of research in modern Astrophysics. While it is very hard to
summarize all the exciting work presented at this meeting, we shall try to
give the reader a glimpse of the state of the art of the field as of April
2006. We shall cite the results by simply the name of the first author on the
corresponding proceedings article, except for papers external to the
proceedings.

\section{The young massive stars in the central parsec}

Young massive ``He-I'' stars dominate the power output of the central
half-parsec \cite{KrabbeEtal95} of our Galaxy. ``Standard'' models of star
formation are not easily applicable here due to a huge tidal field of the
central object at $R= 0.1$ pc distances from \sgra.  The required gas density
is $n_H > 10^{11} \hbox{cm}^{-3} (R/0.1\hbox{pc})^{-3}$.  In the last several
years, a significant breakthrough in understanding of these stars has taken
place. We now know that ...

\begin{itemize}

\item These young stars are not dynamically relaxed (Levin \& Beloborodov
  2003). They appear to form one (J. Lu, A. Ghez) or arguably two stellar
  disks (T. Paumard and also Genzel et al. 2003).

\item There are inner and outer radii that seem to encompass most of the
  massive early type stars, $R\sim 0.03$ and $R\sim 0.5$ pc, respectively.
  (The "S" stars inside of 0.03 pc are discussed separately below.)

\item This implies that the stars are very unlikely to have migrated from
  outside, thus suggesting an {\it in situ} star formation, despite the tidal shear
  produced by Sgr A*.  The previous debate about whether the massive
  stars in the central parsec formed {\it in situ} or whether they are remnant
  of a tidally disrupted cluster brought into the Galactic center by dynamical
  friction is not entirely concluded, but the latter hypothesis seems very 
  strongly constrained, requiring a rather heavy burden of extreme assumptions.  

\item The young stars surrounding Sgr A* are then likely to be the first 
  observational evidence for star formation in massive gaseous disks in 
  the inner parsecs of galactic nuclei (Levin, Paumard).  Recurring {\it in situ}
  star formation was suggested to occur in the GC due to a periodic inflow of
  gas from the Circumnuclear Disk (inner few parsecs) by Morris et al.
  (1999).

\item The inner radius might be indicative of the minimum distance at which
  star formation can still occur, and is consistent with theoretical
  predictions (Levin, Nayakshin). 

\item NIR observations and X-ray constraints suggest that the IMF of young stars
  is very top-heavy (Paumard, Nayakshin). Such an IMF might be the result of
  inefficient cooling in the optically thick conditions in the disk (Nayakshin, Levin).
  The dynamically hot condition of the disk ({\it i.e.}, likely strong turbulence,
  as evidenced by the current circumnuclear disk), and the presumably
  strong magnetic field might also have this effect on the IMF (Morris 1993).  

\item The exact orbits of the stars in the stellar ring(s) remain a
  subject of debate. The acceleration limit (J. Lu) places lower
  limits on the eccentricities of some of the young stars, some of
  which are rather high. This spells trouble for the simplest version
  of the {\it in situ} model, i.e., the model of a gaseous disk in
  circular rotation about \sgra.

\item Numerical simulations, however, show that stars may also form via
  gravitational fragmentation in an eccentric ring of gas (Sanders 1998;
  Nayakshin). Interestingly, an initially flat eccentric disk of stars becomes
  geometrically thick much faster than a flat circular disk (Nayakshin et
  al. 2006) due to orbital precession (Touma). Thus, the fact that the second
  feature in the stellar velocity distribution of young stars in the GC (the
  counter-clockwise ``disk'') is more geometrically thick than the clockwise
  disk may be naturally explained by the different rates of orbital
  precession.


\item New observations constrain stellar effective temperatures with
  increasing reliability. There appear to be no stars more massive
  than about 60 $\msun$, which would be consistent with the age of the
  \sgra\ young star cluster (F. Martins).

\item The star cluster IRS13E is a mysterious collection of several very
  massive young stars and perhaps a couple dozen lighter ones, so it may be
  the surviving core of a massive cluster which has been almost entirely
  tidally stripped (T. Paumard).  Indeed, its dynamics place it in the
  counter-rotating stellar disk, so it may provide evidence that that disk has
  resulted from the disruption of a dense cluster.  However, the visible mass
  of IRS13E is insufficient to bind it gravitationally.  An alternative to the
  rather implausible explanation that the cluster core is only now making its
  first pass near the central black hole is that it is bound by an
  intermediate mass black hole.  However, the X-ray evidence favors colliding
  stellar winds over black hole accretion. IRS13E-like objects are in fact not
  formed naturally in current {\it in situ} star formation simulations
  (Nayakshin). Understanding IRS13E is therefore crucial for understanding of
  the overall star formation event in the GC.

\item Other co-moving groups of young stars might be present (R. Schoedel) in
  the inner parsec.

\item IRS16SW appears to be an eclipsing contact binary with two $\sim
50~\msun$ stars orbiting with a period of about 18 days (T. Ott,
M. Rafelski, F. Martins et al.\ 2006).

\end{itemize}

\section{S-stars in the central $\sim 0.03$ parsec}

The so-called ''S-stars'' are the stars observed in the inner arcsecond in
projection from Sgr~A*, i.e., in the inner region that excludes the He-I stars
discussed above. These stars are now spectroscopically resolved to be less
massive but nonetheless still quite young B-type stars. The paradox of youth
is orders of magnitude more severe for these stars than for the He-I stars.
Furthermore, the orientations of the orbits of these stars appear to be
consistent with isotropic, so they do not point to any organized formation
mechanism such as a disk.  A large fraction of the S-stars has relatively
eccentric orbits, although this fraction is difficult to determine reliably
because of selection effects.  There is also a strong selection effect toward
studying only the brightest sources, although more stars are emerging all the
time, so this situation is improving.

While there were many theoretical suggestions for the origin of these stars, 
there does not yet appear to be a clear winner among them.  Gravitational 
instability of a massive accretion disk is not predicted to form stars closer 
in than about $1''$, and hence the in-situ star formation for S-stars is not 
favored. Some of the new ideas emerged from the meeting on the origin 
of these stars are:

\begin{itemize}

\item Massive perturbers, such as massive molecular clouds or star clusters,
  can significantly enhance the rate at which massive young binaries diffuse
  on nearly radial orbits (H. Perets). If these binaries are then disrupted, a
  population of stars similar to S-stars might be created.

\item The population of remnants of disrupted binaries would have very large
  eccentricities ($e\approx 0.99$; Y. Levin). Ordinary N-body relaxation is
  too slow to evolve the orbits, which would then be a serious problem for the
  model, as several of the S-stars have moderated eccentricities ($e <
  0.5$). Resonant relaxation (C. Hopman, T. Alexander), might however result
  in the orbital evolution that is rapid enough to explain those ``rouge''
  S-stars.

\item Although this was not discussed at the meeting, we note that the
  relatively young, ultra-high velocity stars now being found in the galactic
  halo, whose orbits are consistent with being purely radial, are widely seen
  to result from binary encounters either with the central black hole or with
  stellar mass black holes in a central cluster.  Such process may be a
  promising way of producing the S-stars as the former companions of the
  ultra-high velocity stars (O'Leary \& Loeb 2006).

\item Spectra of the S-stars are consistent with those of normal B-type stars
  (F. Martins), possibly ruling out the model in which the S-stars are created by
  tidal disruption of giants on nearly radial orbits.

\end{itemize}

\section{NIR/X-ray flares from \sgra}

At the time of the last workshop on the Galactic Center, \sgra\ had
not yet been discovered at infrared wavelengths, and the first
observations of X-ray flares had only recently been announced
(Baganoff et al. 2001).  Millimeter-wavelength flares had been
reported already at the 1998 Tucson workshop, and continued follow-up
work on that was reported here by A. Miyazaki. Since that time, the
characteristics of the emission have become clearer, and theories for
the spectrum of \sgra\ have been elaborated to account for that
(F. Yuan).  However, this study is still at a very early stage, both
observationally and theoretically, as the frequency, spectrum,
location, and cause of the flares all remain widely debated.  X-ray
flares are well-defined events, rising above the feeble, marginally
extended ($\sim$ 1.5") quiescent emission almost once per day.  The
near-infrared intensity variations of more than an order of magnitude
appear to be essentially continuous red noise variations, rather than
being well-defined "flares".  The common use of the term "flare" in
the NIR typically refers to a broad intensity maximum in the
continuous light curve.

\begin{itemize}

\item The locations of NIR flares are offset from the dynamical center of the 
  \sgra\ star cluster by no more than 2 mas $\sim 2 \times 10^{14}$ cm. This 
  implies they do originate from \sgra\ (S. Trippe).

\item There is a dusty clump of gas very near the line of sight to
  \sgra, which may influence the mid-infrared fluxes of \sgra\ during flares 
  (Ghez et al.\ 2005; A. Eckart).  
  
\item There is not yet universal agreement about whether the spectral index 
  observed during flares is correlated with the flare luminosity, but much recent
  effort has gone into measuring the NIR spectrum and its relationship to the
  X-ray emission properties (S. Hornstein;  S. Gillessen; A. Krabbe; A. Eckart).
  Hornstein finds no variability of the NIR spectral index of \sgra\ from 1.6 to 4.6 
  $\mu$m.  

\item Simultaneous observations of \sgra\ at infrared, radio and X-ray
  wavelengths have been attempted since 2002, and a few events have been
  captured in multiple bands (A. Eckart; Yusef-Zadeh et al. 2006).  The
  present indications are that X-ray flares are accompanied by NIR maxima,
  although the statistics on this can only be improved.  This strengthens
  theories in which the same population of electrons is responsible for both
  the NIR and X-ray events, but probably by virtue of different processes
  (synchrotron and synchrotron self-Compton, respectively).

\item A breakthrough has occurred with the first measurements of the 
  polarization of the NIR emission from \sgra and of its variability (A. Eckart).
  The polarization shows much more time structure than the total intensity,
  suggesting that it is affected by the varying direction of motion of the 
  nonthermally emitting gas.  A similar phenomenon is also evidenced at
  submillimeter wavelengths (D. Marrone).  Polarization variability time scales 
  of $\sim$ 20 minutes are particularly intriguing in view of the occasional 
  periodicities that have been claimed for the total IR (Genzel et al. 2003) 
  and X-ray intensities (G. Belanger), and in view of the coincidence of this 
  time scale with the period of the innermost stable circular orbit around a
  rotating black hole of $3.6~\times~10^6~\msun$.  Theoretical models 
  suggest that polarised emission of a rotating blob near \sgra\ could 
  produce characteristic signatures that would allow
  observers to constrain the spin of \sgra (L. Meyer; A. Broderick).
  
\item   A compelling 22-minute periodicity was inferred from the data on 
  one rather long X-ray flare observed with XMM (G. Belanger).  The 
  duration of the entire flare was only about 7 times this period, so the 
  analysis of the statistics of this event is an important and delicate issue.
  The $\sim$15 other flares seen with Chandra and XMM have not yet 
  shown any significant periodicity, so any such modulation of the X-ray
  flare signal is apparently uncommon.

\end{itemize}

\section{New X-ray results on the GC}

\begin{itemize}

\item The {\em Suzaku} telescope, formerly known as {\em AstroE2}, can
 now distinguish between rather closely lying X-ray lines. K. Koyama
 reported {\em Suzaku} results on the Fe K-$alpha$ line at $\sim 6.7$
 keV in the GC diffuse emission. It is found that the centroid energy
 is 6.680 keV, which favors the thermal collisional origin of that
 line, rather than its being due to charge exchange with cosmic rays
 (in which the line energy would be 6.666 keV). The ratio of X-ray
 line intensities shows that plasma emitting the diffuse emission is
 in near ionization equilibrium.  Also, the emitting medium does not
 have a large bulk motion with respect to \sgra.

\item The diffuse X-ray emission presented a puzzle for many years, as the
  plasma emitting it would be too hot to be confined by the gravity of the GC,
  and would require an enormous energy source to maintain (R. Belmont).
  R. Warwick suggests that the diffuse X-ray emission is produced by a
  population of cataclysmic variables (CVs) rather than by a hot diffuse
  plasma.  This parallels the recent results of Revnivtsev et al. (2006) on
  the larger scale Galactic ridge X-ray emission, but conflicts with the
  conclusions of Muno et al. (2004) and Koyama et al. (2006).

\item Association of the ``neutral'' 6.4 keV Fe K-$\alpha$ line with the
  $4.5-6$ keV continuum emission argues for origin of most of that emission in
  local cosmic ray ionization (R. Warwick), rather than photo-ionization.

\item Sgr B2 cloud's 6.4 keV Fe K-$\alpha$ X-ray emission, and that in the
  newly found ``X-ray reflection nebula'', are however the results of
  photo-ionization (K. Koyama).

\item {\em Chandra} observations show that low mass stars $M \simlt
  1-2 \msun$ seem to be deficient by a factor of up to ten in both the
  central star cluster (S. Nayakshin) and the Arches star cluster
  (D. Wang).  The extent to which the apparently top-heavy MF of the
  Arches cluster is owed to the preferential tidal stripping of its
  original low mass stars is not yet clear.

\end{itemize}

\section{The central molecular zone and star formation}

\begin{itemize}

\item Gas in the CMZ is particularly warm, dense, and turbulent, with temperatures
  ranging up to several hundred degrees K (N. Rodriguez-Fernandez).  About 10-30\%
  of the column density of the CMZ can be attributed to a hot ($\sim$150 K) 
  photodissociation region.  It is not yet possible to decide definitively among 
  the possible heating mechanisms proposed for the rest of the gas, or even 
  whether the right mechanism has yet been proposed.  Observations of an
  unusually high relative abundance of SiO in CMZ clouds imply that shocks 
  affect the chemical abundances there.  

\item The study of the CMZ is a mature field, with many surveys done in past decades.
  Therefore, it was remarkable to see in a recent CO survey some altogether new 
  morphological features: very large molecular loops rising far above the 
  Galactic plane (Y. Fukui).  The shapes and velocity fields of these features are 
  reasonably interpreted as having been produced by the Parker instability, 
  although it remains unclear why the loop-like magnetic structures resulting from
  that instability would have a molecular form.

\item The Spitzer/IRAC survey has provided a wealth of detail that will be mined for
  a long time to come (S. Stolovy).   Hot dust and PAH emission are ubiquitous
  throughout the inner few hundred parsecs, pointing the way to sites of star
  formation activity.  The close correspondence between the IRAC images
  and the radio images of HII regions -- notably near the Quintuplet Cluster --
  illustrate the strong interaction between the luminous, hot stars and gas in 
  the CMZ (A. Cotera).  

\item Star formation in the CMZ is a currently active process that has been 
  ongoing quasi-continuously since the formation of the Galaxy (Figer 2007).  
  It has been punctuated recently, and very likely at a steady rate in the past,
  by the formation of massive, compact clusters, three of which have formed
  within the past 10$^7$ years.  Inevitable tidal dissolution makes it very
  difficult to identify older clusters than that.  One of the important outstanding
  problems in Galactic center research is to identify the mechanism for the
  formation of such massive clusters as the Arches and Quintuplet.   

\end{itemize}

\section{The Galactic center magnetic field}

Two important puzzles that emerge from this workshop are the strength 
and the dispersion of the magnetic field in the central 100 parsecs (M. Morris).  
The paradigm suggested some time ago by the presence of numerous,
vertically-oriented, nonthermal radio filaments was a pervasive milligauss
field having a relatively small dispersion in both strength and direction.   
The field was hypothesized to be highlighted by synchrotron radiation 
only at the places where relativistic electrons have been injected.  The
pervasive milligauss field has been called into question based on recent 
studies of low-frequency synchrotron radiation (T. LaRosa),
and using the minimum energy assumption, which may be inapplicable to
this situation.   The debate about the field strength hinges also on the
relatively short synchrotron lifetime of the emitting electrons in the strong
field case, and therefore whether there exists a reacceleration mechanism.
Others have argued that the radio filaments represent
local concentrations of the field caused by turbulence or by some transitory
current structure.  However, these proposals encounter difficulties with time
scales or geometry.  Additional constraints on the magnetic field can perhaps
be supplied by the high energy results from INTEGRAL (G. Belanger) and 
HESS (J. Hinton).   What is now needed is a direct measurement of the 
magnetic field; 
submillimeter measurements of polarized line and continuum emission 
may settle this issue before the next workshop.  Also, current theoretical
efforts to model cosmic rays and their synchrotron emission will soon be
informing this debate. It is an important issue to settle, because a strong
magnetic field could play a vital role in many of the phenomena observed
in the Galactic center.

\section{Other interesting results}

\begin{itemize}

\item Adaptive optics near-infrared observations of the Arches cluster
  also suggest that its IMF is top-heavy (A. Stolte, S. Kim). The IMF
  is mass-segregated as a function of the projected distance from the
  center of the cluster.
  
\item  The prominent mid-infrared source AFGL 5376 appears to have 
  resulted from a strong, $\sim$100-pc shock between two vertically oriented 
  molecular systems.  The association of this structure with magnetic radio
  filaments (J. Staguhn) raises the possibility that the shock is associated 
  with a region of strongly compressed magnetic field.  

\item The Sgr A East supernova remnant, which encircles the GC in projection,
  and is probably not very far behind \sgra\, is impacting the dense molecular
  material surrounding it, giving rise to 1720 MHz OH maser emission 
  (L. Sjouwerman) and to shocked H$_2$ line emission (S. Lee) at several
  points around its periphery.  The impact upon the 50 km s$^{-1}$ 
  cloud gives rise to a compressed ridge (M. Tsuboi).  The energetics and the age
  of this supernova are still under discussion.  While some consider it to be a normal
  Type II supernova, Others suggest that it may have resulted from a much more 
  energetic hypernova.  Similarly, age estimates for the remnant now range between
  1700 and 10,000 years. 
  
\item The mass of the Circumnuclear Disk (CND) , which orbits \sgra\ and has an 
  inner radius of $\sim$1 pc, has been debated for some time, with estimates
  ranging from $10^4$ to a few $\times~10^5~\msun$.   With the SMA results
  of M. Montero-Castano on J = 4-3 HCN emission, it now appears that, while
  hot gas is characteristic of the CND, the density is not uniformly as high as 
  had previously been estimated (Christopher et al. 2005).   Consequently, the 
  CND mass may lie near the lower end of the above range.  The CND
  is a likely reservoir for future accretion activity onto SgrA*, so this result has 
  important implications for the long-term accretion rate.  The CND also serves
  as a test case for the clumpy medium model of T. Beckert for dusty tori in 
  low-luminosity AGNs.

\item There appears to be a low extinction channel along the Galactic
  plane, from NE to SW which points to the minicavity (R. Sch\"odel).
  If ascribed to an outflow from \sgra\, it agrees with the model of
  K. Mu\v{z}i\'{c} for the proper motions of mid-IR filaments, and
  with the orientation of the polarization vector found by A. Eckart.

\item The GC distance determined using the orbital fits to the proper motions of
  the S-stars, in concert with measurements of their radial velocities, could be as
  small as 7.4 kiloparsec (T. Ott), although ongoing efforts are warranted to 
  eliminate astrometric biases occurring near periapse and during close passages
  by other stars.  

\item Sophisticated Monte-Carlo modeling of dynamical mass segregation in the
  GC predicts about 20 thousand stellar-mass black holes in the inner parsec
  (M. Freitag).

\item Horizontal branch/Red clump stars, i.e. stars with $M\simlt 2 \msun$ and
  older than $10^9$ years, seem to be missing in the inner 9'', possibly due
  to their expulsion from that region by stellar mass black holes
  (T. Alexander).

\item Numerical models of stellar wind accretion onto \sgra\ predict a strong
  variability of the accretion rate (J. Cuadra) on time scales of tens to
  hundreds of years. The implications of these results for the structure of the
  inner radiatively inefficient accretion flow (F. Yuan) are at the moment not
  clear.  Radio evidence gathered over a long term with the VLA indicates
  that there has been a secular change in the low-frequency turnover 
  frequency (T. An), suggesting a variable mass flux in the stellar winds which 
  feed into the accretion flow.  These variations would modulate the opacity
  of the local absorbing medium.  

\item Moving toward higher frequencies so that the $\lambda^2$
  dependence of the size of the elliptical scattering disk is
  minimized, several VLBI groups are reporting an intrinsic size of
  for the radio source, as well as its variation with wavelength
  (G. Bower, Z-Q. Shen, T. Krichbaum).  The intrinsic size at 3mm is
  $\sim$15 Schwarzschild radii, consistent with equating the dominant
  variability time scales reported at that wavelength with the
  dynamical time at that radius (Mauerhan et al. 2005).  There is
  still some discrepancy between the suggested wavelength-dependences
  of the intrinsic source size: $\lambda^{1.1}$ (Shen) to
  $\lambda^{1.6}$ (Bower).
  
  There is enthusiasm in this community for the future use of VLBI techniques 
  at submillimeter wavelengths, where the scattering disk will presumably be 
  substantially smaller than the source.  In one of the next few meetings, we
  can perhaps expect to witness images of the shadow of the black hole's
  event horizon.  As A. Broderick shows, the theorists are ready for that.  

\end{itemize}

\ack The conference organizers, and especially Rainer Sch\"odel, must
be thanked for organizing this workshop, running it amazingly well and
problem-free, which allowed the participants to concentrate on the
exciting science of the galactic Center.


\end{document}